\documentclass{aa}
\usepackage{graphicx}
\usepackage{amssymb}
\usepackage{natbib}
\bibpunct{(}{)}{;}{a}{}{,}  

\newcommand{\eqb}{\begin{eqnarray}}
\newcommand{\eqe}{\end{eqnarray}}
\newcommand{\sth}{\sigma_{\rm T}}

\newcommand{\sgg}{\sigma_{\gamma \gamma}}

\newcommand{\nelec}{n_{\rm e}}
\newcommand{\nphot}{n_{\gamma}}
\newcommand{\melec}{m_{\rm e}}

\newcommand{\ggabs}{{\gamma\gamma}}
\newcommand{\eg}{\epsilon_\gamma}
\newcommand{\egmn}{\epsilon_{{\gamma}{\rm min}}}
\newcommand{\egmx}{\epsilon_{{\gamma}{\rm max}}}
\newcommand{\lginj}{\ell_\gamma^{\rm inj}}
\newcommand{\lginjcr}{\ell_{\gamma,{\rm cr}}^{\rm inj}}
\newcommand{\lgh}{\ell_\gamma^{h}}
\newcommand{\lgs}{\ell_\gamma^{\rm s}}
\newcommand{\lgo}{\ell_\gamma^{\rm out}}
\newcommand{\numn}{\nu_{\rm min}^{\rm obs}}
\newcommand{\numx}{\nu_{\rm max}^{\rm obs}}
\newcommand{\bquemx}{B_{\rm q,mx}}
\newcommand{\tgesc}{t_{\gamma,\rm esc}}
\newcommand{\teesc}{t_{\rm e,esc}}
\newcommand{\egmneff}{\epsilon_{\gamma{\rm min,eff}}}
\newcommand{\esynmx}{x_{\rm max}}
\newcommand{\egamma}{\epsilon_{\gamma}}
\newcommand{\linj}{l_{\gamma}^{\rm inj}}
\newcommand{\lcr}{l_{\gamma, \rm cr}^{\rm inj}}
\newcommand{\lB}{l_{\rm B}}
\newcommand{\gcr}{\gamma_{\rm cr}}
\newcommand{\gmax}{\gamma_{\rm max}}
\newcommand{\xmax}{x_{\rm max}}
\begin{document}

\title{Implications of automatic photon quenching
on compact gamma-ray sources}
\author{M. Petropoulou \and A. Mastichiadis}
\institute{Department of Physics, University of Athens, Panepistimiopolis, GR 15783 Zografos, Greece}
\offprints{A. Mastichiadis}
\date{Received ... / Accepted ...}
\abstract{}{ 
We investigate photon quenching in compact 
non-thermal sources.
This involves photon-photon annihilation
and lepton synchrotron radiation in a network that
can become non-linear. As a result
the $\gamma-$ray luminosity of a source
cannot exceed a critical limit that 
depends only on the radius of the source
and on the magnetic field.
}
{We perform analytic and numerical calculations that
verify previous results and extend them so that the
basic properties of photon quenching are investigated.
}
{We apply the above to the 2006 TeV observations of 
quasar 3C279 and obtain
the parameter space of allowed values for the radius of 
the emitting source,
its magnetic field strength and the Doppler factor of the flow.
We argue that the TeV observations favour either a modest
Doppler factor and a low magnetic field or a high Doppler
factor and a high magnetic field.
}
{}

\titlerunning{Automatic $\gamma$-ray quenching in compact sources}
\authorrunning{Petropoulou \& Mastichiadis}
\maketitle

\section{Introduction}

Astrophysical $\gamma-$ray sources are often subject to the so-called 
`compactness' problem: when the flux at longer wavelengths
is combined with the dimensions of the source, which are inferred from light-crossing time arguments, 
 a high optical thickness
to photon-photon annihilation and effective absorption of $\gamma-$rays are implied.

The possibility that high-energy photons could
pair-produce on soft target photons instead of escape in compact sources
was first discussed by \cite{jelley66}, while the ratio of the luminosity
to the size of the source, $L/R$, emerges as a determining factor of
whether or not a high-energy photon will actually be absorbed \citep{herterich74}.
 This was followed by work that
took into account photon-photon annihilation not only as a sink
of $\gamma-$rays, but also as a source of electron-positron pairs inside
non-thermal compact sources 
\citep{bonometto71, guilbert83,
kazanas84, zdziarski85, svensson87}. 
The aim of these models was to calculate self-consistently the 
photon flux escaping from the sources by taking into account 
the energy redistribution caused by  
the $\gamma-$annihilation-induced  pair cascades. The assumption was 
that high-energy particles or $\gamma-$rays were injected
uniformly in a source that also contains soft photons, and the 
system was followed to its final steady-state through the
solution of a set of kinetic equations describing the physical 
processes at work.
Several groups have also developed numerical codes 
for the computation of time-dependent
solutions to the kinetic equations taking photon-photon annihilation
into account \citep{coppi92, mastkirk95, stern95, boettcherchiang02}. 
These algorithms are commonly used in source modelling 
\citep{mastkirk97, kataoka00, konopelko03, katarz05}. 

One of the assumptions of these models, as stated above, is  the
presence of soft photons in the source, which serve as targets
for the $\gamma-$ray annihilation.  
A different approach has been recently presented by \cite{SK07}, henceforth SK.
These authors focused on the non-linear effects induced by
photon-photon annihilation and investigated the necessary conditions 
under which $\gamma$-ray photons can cause runaway pair production\footnote{
We note that in the same framework an analogous study where ultrarelativistic protons are
the primary injected particles has earlier been presented by \cite{kirkmast92}.}. 
\cite{SK07} showed that  there is a limit to the $\gamma-$ray luminosity
escaping from a source, which
does not rely on the existing soft photon population 
but is instead a theoretical limit depending 
only on parameters such as the source
size and its magnetic field strength. Violation of this limit
leads to 
automatic quenching of the $\gamma-$rays.
This involves a network of processes, namely photon-photon
annihilation and lepton synchrotron radiation, which can
become non-linear once certain criteria are satisfied.
In this case electron-positron pairs grow 
spontaneously in the system and the `excessive'
$\gamma-$rays are absorbed
on the synchrotron photons emitted by the pairs. As a result
the system reaches a final steady state where the 
$\gamma-$rays, soft photons, and electron-positron pairs
have all reached equilibrium. Therefore this 
can occur even in the hypothetical case when
there are no soft photons in the source, at least
initially. 

Indeed, 
photon quenching has some interesting implications for
emission models of $\gamma-$rays because it 
gives a robust 
upper limit on the $\gamma-$ray luminosity. The
aim of the present paper is to investigate the implications
of this network for the high - energy compact
astrophysical sources. 
In \S2 we will expand the analytical 
approach of SK, who studied 
the dynamical system of photons and relativistic pairs
 using $\delta-$functions for the photon-photon annihilation cross section
and for the synchrotron emissivity, while they treated synchrotron losses as catastrophic.
In \S3 we will use a numerical approach that will allow us to
study the properties of quenching using the full cross section for $\gamma\gamma$ absorption
and the full synchrotron emissivity. As an example, in \S4
the above will be applied to the 2006 MAGIC TeV observations
of quasar 3C279 to extract a parameter space of allowed values for the source parameters, i.e,
the radius $R$ and the magnetic field strength $B$, as well as for the Doppler factor $\delta$ of the flow.
Finally in \S5 we conclude and give a brief discussion of the main points
of the present work.

\section{Analytical approach}

\subsection{First principles}
We begin by assuming a spherical source of radius $R$ containing
a tangled magnetic field of strength $B$. We also assume that
monoenergetic $\gamma-$ rays of energy $\eg$ (in units of 
$\melec c^2$) are uniformly produced by some unspecified 
mechanism throughout the volume of the source. If these are
injected with a luminosity $L_\gamma^{\rm inj}$,  one
can define the injected $\gamma-$ray compactness as
\eqb 
\label{lgg}
\lginj={{L_\gamma^{\rm inj}\sigma_T}\over
{4\pi R\melec c^3}},
\eqe
where $\sigma_T$ is the Thomson cross section. 
Without any substantial soft photon population 
inside the source, the $\gamma-$rays
will escape without any attenuation in one crossing time.
However, as SK showed, the injected 
$\gamma-$ray compactness cannot 
become arbitrarily high because if a critical value is reached, 
the following loop starts operating

1. Gamma-rays pair-produce on soft photons, which can be
arbitrarily low inside the source.

2. The produced electron-positron pairs cool by emitting synchrotron
photons, thus acting as a source of soft photons.

3. The soft photons serve as targets for more $\ggabs$ interactions.

There are two conditions that should
be satisfied simultaneously for this 
network to occur: The first, which is a {\sl{feedback}} condition,
requires that the synchrotron photons emitted from the pairs
have sufficient energy to pair-produce on the $\gamma-$rays.
By making suitable simplifying assumptions, one can derive 
an analytic relation for it -- see also SK. Thus, 
combining (1) the threshold condition for $\ggabs$-absorption
$\eg\epsilon_0=2$, 
(2) the fact that there is equipartition of energy among
the created electron-positron pairs $\gamma_p=\gamma_e=\gamma=\eg/2$
and (3) the assumption that the required
soft photons of energy $\epsilon_0$
are the synchrotron photons that the electrons/positrons radiate,
i.e., $\epsilon_0=b\gamma^2$
where $b=B/B_{\rm crit}$ and $B_{\rm crit}=(\melec^2 c^3)/(e\hbar)
\simeq 4.4\times 10^{13}$~G is the critical value of the magnetic field,
one derives the minimum value of the magnetic field required
for quenching to become relevant
\eqb
B_q=8\eg^{-3}B_{\rm{crit}}.
\label{bcrit}
\eqe
Thus for $B\ge B_q$ the feedback criterion is satisfied.
Note that this is a condition that only contains the emitted 
$\gamma-$ray energy and the magnetic field strength;
moreover, it can easily be satisfied,
at least if one is to use the values inferred from typical
modelling of the sources \citep{boettcher07, boettcher09}.

Along with the feedback criterion, a second criterion must
be imposed for the quenching to be fully operative. 
A simple way to see this is with the following consideration.
Assume that the $\gamma-$rays
pair-produce on some soft photon and that the 
created electron-positron pairs
cool by emitting synchrotron photons. Because an electron
emits several such photons before cooling the critical condition
occurs if the number density of the $\gamma-$rays is such that
at least one of the synchrotron  photons pair-produces on
a $\gamma-$ray instead of escaping from the source. Thus the condition
for criticality can be written as
\eqb
{\cal{N}}_{\rm s}(\eg/2) n(\eg) \sigma_\ggabs R\ge 1,
\eqe      
where $n(\eg)$ is the number density of $\gamma-$rays, $\sigma_\ggabs$
is the cross section for $\ggabs$ interactions and $\cal{N}_{\rm s}(\gamma)$
is the number of synchrotron photons emitted by an electron with Lorentz factor $\gamma$ 
before it cools. Assuming that the pair-producing collisions occur
close to threshold, approximating the cross section there by
$\sigma_\ggabs\simeq\sigma_T/3$ and using 
${\cal{N}}_{\rm s}(\gamma)\simeq\frac{\gamma}{b\gamma^2}$ and
 $n(\eg)=\frac{L_{\gamma}^{\rm inj}}{V\eg \melec c^2}\frac{R}{c}$, the critical condition 
can be written as
\eqb
\label{lcr0}
\eg\lginj\ge 4,
\eqe
where the feedback condition (\ref{bcrit}) and eq. (\ref{lgg}) were also used. 
Taken as equalities, relations (\ref{bcrit}) and (\ref{lcr0}) define the
{\sl{marginal stability}} criterion, which essentially is a
condition for the $\gamma-$ray luminosity.

\subsection{Kinetic equations}

We begin by writing 
the kinetic equations that describe the distributions of $\gamma$-ray
photons, soft photons, and electrons in the source.
These are respectively
\eqb
 \label{gammaray} \frac{\partial n(\egamma,\tau)}{\partial \tau}
+n= \cal{Q}^{\gamma}_{\rm inj}+\cal{L}^{\gamma}_{\gamma \gamma}
  \eqe

\eqb \label{soft} \frac{\partial n_0(x,\tau)}{\partial \tau} +n_0 =
\cal{Q}^{\rm s}_{\rm syn}
  \eqe

and 
\eqb \label{elec} \frac{\partial n_e(\gamma,\tau)}{\partial
\tau} = \cal{Q}^{\rm e}_{\gamma \gamma}+\cal{L}^{\rm e}_{\rm syn},
  \eqe
where $n$, $n_0$ and, $n_e$ are the differential $\gamma$-ray, soft photon, and electron
  number densities, respectively, and 
 $\egamma$, $x$, and $\gamma$ are the corresponding energies normalized
  in $\melec c^2$ units. The densities refer to the number
  of particles contained in a volume element $\sth R$. In other
  words, if $\hat{n}_i$ expresses the number of particles of species
  $i$ per ergs per cm$^3$, then $n_i = \hat{n}_i(\sth R)(\melec c^2)$.
  Time is normalized with respect to the photon crossing/escape time from 
the
  source, $t_{\rm cr}=R/c$. Thus, $\tau$, which appears in the kinetic equations, is
  dimensionless and equals $\tau=\frac{ct}{R}$. 
The operators $\cal{Q}$ and $\cal{L}$ denote injection
and losses, respectively.
 The only
processes that we take into account are $\gamma \gamma$ annihilation 
and synchrotron cooling of the produced pairs. The
synchrotron emissivity is approximated by a $\delta$-function, i.e.,
$j_s(x)=j_0 \delta(x-\epsilon_0)$, where $\epsilon_0=b\gamma^2$ is
 the synchrotron critical energy. In all cases we will assume 
a monoenergetic $\gamma-$ray injection at energy $\egamma$.

\subsubsection{$\delta$ - function approximation for the $\sgg$ cross section}
First we will examine the stability of the system using the simplest form
for the photon-photon annihilation cross section \citep{zdziarski85}:
\eqb
\label{s_delta}
\sigma_{\gamma \gamma}(x)=\frac{\sth}{3} x
 \delta \left(x-\frac{2}{\egamma}\right),
\eqe
where $x,\eg$ are the energies of the soft and $\gamma$-ray photons, respectively.

The particle injection and loss operators take the following forms:
  \eqb
  \label{operators}
   \cal{Q}^{\gamma}_{\rm inj} & = & \frac{3 \linj}{\egamma^2} \\
  \cal{Q}^{\rm s}_{\rm syn} & = & \frac{2}{3} \lB b^{-3/2} x^{-1/2}
  n_e(\sqrt{x/b},\tau) \\
  \cal{L}^{\gamma}_{\gamma \gamma} & = & -\frac{2}{3}\frac{n_0\left(2/\egamma\right) n(\egamma)}{\egamma}\delta(\eg-2\gamma)\\
  \cal{L}^{\rm e}_{\rm syn} & = & +\frac{4}{3}\lB \frac{\partial}{\partial
  \gamma}\left(\gamma^2 n_e\right)
  \eqe
  and
  \eqb
  \cal{Q}^{\rm e} _{\gamma \gamma} & =& -4 \cal{L}^{\gamma}_{\gamma \gamma},
  \eqe
  because each photon-photon annihilation results in a pair of leptons,
  each one with approximately half of the initial $\gamma$-ray
  energy. In the above equations the `magnetic compactness' was introduced
  \eqb
  \lB= \left(\frac{U_{\rm B}}{\melec c^2}\right)\sth R \cdot
  \eqe
The system of eq. (\ref{gammaray})-(\ref{elec}) can now be written
as
 \eqb
 \label{systemfull}
 \frac{\partial n(\egamma)}{\partial \tau} &=& -n(\egamma)-\frac{2}{3}\frac{n_0\left(2/\egamma\right)
 n(\egamma)}{\egamma}\delta(\eg-2\gamma) + \frac{3 \linj}{\egamma^2} \nonumber \\
 \frac{\partial n_0(x)}{\partial \tau} & = & -n_0(x) + \frac{2}{3} \lB b^{-3/2} x^{-1/2}
  n_e \left(\sqrt{x/b}\right)\nonumber \\
  \frac{\partial n_e(\gamma)}{\partial \tau} & = & +\frac{4}{3}\frac{n_0\left(2/\egamma \right) n(\egamma)}{\egamma}
   \delta\left(\gamma-\eg/2\right)+ \nonumber \\
  & +& \frac{4}{3}\lB\frac{\partial}{\partial\gamma}\left(\gamma^2 n_e(\gamma)\right) \cdot
 \eqe
In this section we will examine the stability of the trivial
stationary solution of the system (\ref{systemfull}):
\eqb
\label{trivial} 
\begin{array}{l l l}
\bar{n} = 3 \linj/\egamma^2, & \bar{n}_0 =  0, & \bar{n}_e =  0,
\end{array}
 \eqe

 which corresponds
to the free propagation of $\gamma$-ray photons through the source.
To investigate the stability of the system, we assume that
initially arbitrarily small perturbations of the soft photon $n_0'$ and electron
$n_e'$ densities are present, which lead to the perturbation of the 
$\gamma$-ray photon density $n'$. After linearization, the system
(\ref{systemfull}) becomes
 \eqb
 \label{systemlinear}
\frac{\partial n'}{\partial \tau} & = & -n'-\frac{2}{3}\frac{n_0'\bar{n}}{\egamma}\delta(\eg-2\gamma) \nonumber \\
\frac{\partial n_0'}{\partial \tau} & = & -n_0'+ \frac{2}{3} \lB
b^{-3/2} x^{-1/2}n_e' \nonumber \\
\frac{\partial n_e'}{\partial \tau} & = & +\frac{4}{3}\frac{n_0'
\bar{n}}{\egamma}\delta \left(\gamma-\eg/2\right)+\frac{4}{3}\lB \frac{\partial}{\partial
  \gamma}\left(\gamma^2 n_e'\right) \cdot
 \eqe
The stability analysis of the system above can be simplified when
one works with the Laplace transformed number densities:
 \eqb
 n_0'(x,\tau) \rightarrow n_0'(x,s)= \int_0^{\infty} d\tau \
 n_0'(x,\tau)e^{-s\tau},
 \eqe
with $s$ being the solution to the eigenvalue problem:
\eqb
\label{hard}
(s+1) n'(\egamma,s) & = & -\frac{2}{3}\frac{n_0'\left(2/\egamma, s\right) \bar{n}}{\egamma} \eqe
\eqb
\label{photons}
 (s+1)n_0'(x,s) & = & \frac{2}{3} \lB b^{-3/2} x^{-1/2}n_e'(\sqrt{x/b},s)
 \eqe
 \eqb
\label{electrons}
  s n_e'(\gamma,s) & = &+ \frac{4}{3}\frac{n_0'\left(2/\egamma,s \right) \bar{n}}{\egamma}\delta\left(\gamma-\eg/2\right)+\nonumber \\
& & +\frac{4}{3}\lB\frac{\partial}{\partial
  \gamma}\left(\gamma^2 n_e'(\gamma,s)\right)\cdot
 \eqe
 For large $\tau$ both the soft photon and electron distributions
 behave as $e^{s\tau}$.  If $s=-1$, then $n_0'=n_e'=0$ and the
 $\gamma$-ray photons freely escape from the source. If $s>0$, 
 one finds solutions that grow with elapsing time.
The marginally stable solution, which is obtained when one sets $s=0$, will be examined below.

Combining equations (\ref{hard})-(\ref{electrons}) leads to an ordinary differential equation for the 
perturbed electron distribution with the following solution:
\eqb
\gamma^2 n_e'(\gamma) = C-\frac{2\bar{n}n_e'(\gcr)}{3b^{3/2} \sqrt{2\eg}},
\eqe 
where $\gcr=\sqrt{\frac{2}{b\eg}}$. Setting $\gamma=\gcr$ in the above equation one
finds
\eqb
\label{elec_final}
n_e'(\gamma)=C \left(\frac{\gcr}{\gamma}\right)^2 \frac{1}{\gcr^2+\frac{2\bar{n}}{3b^{3/2}\sqrt{2\eg}}}\cdot
\eqe
Then the perturbed photon distribution is found using eq.(\ref{photons}).

As already mentioned in \S 2.1, the $\gamma$-ray compactness cannot become
arbitrarily high but instead saturates at a critical value $\lcr$. 
When $\linj<<\lcr$ only a negligible fraction of the 
hard photon luminosity will be absorbed, i.e., $l_{\gamma}^{h'} \rightarrow 0$, or equivalently $l_{\gamma}^{h}\propto \linj$. 
 We call critical the injected luminosity required for $l_{\gamma}^{h'}=l_{\gamma}^{s'}$ to occur. After 
this point
 any further increase of the
injected luminosity will be seen as an increase of the luminosity in soft photons.
Using eq.(\ref{hard}), (\ref{photons}), and (\ref{elec_final}) one derives
\eqb
 \egamma^2n'(\eg;\lcr) & = & \int^{\xmax} dx \ x n_0'(x;\lcr) ,
\eqe
where $\xmax=b\eg^2/4$. Solving the above equation with respect to $\lcr$, we find
\eqb
\label{lcrit_delta}
\lcr=\sqrt{2 b\eg}.
\eqe 
\begin{figure}[h!]
 \centering
\includegraphics[width=8.5cm, height=8cm]{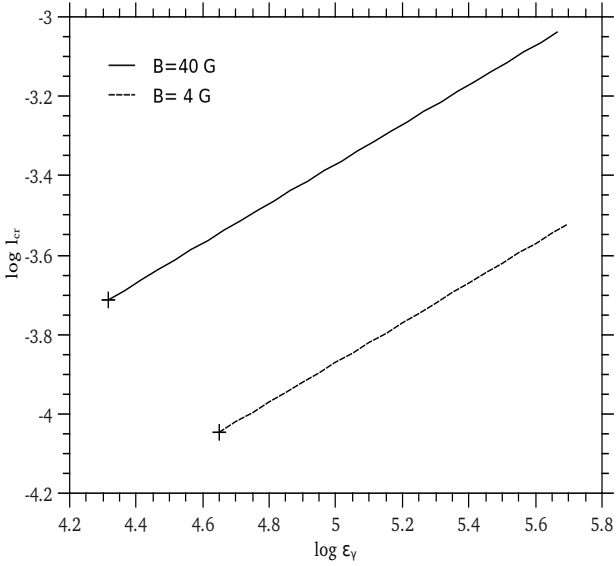}
\caption{Log-log plot of the critical luminosity compactness $\lcr$ as a function of the injected $\gamma$-
ray photon energy $\egamma$, when the $\delta$-function approximation for the cross section $\sgg$ was used for two
values of the magnetic field strength: $B=40$ G (solid line) and $B=4$ G (dashed line). 
The values calculated by SK using only catastrophic losses are marked with crosses.}
\label{lcr1}
\end{figure}

To compare 
our results with the value $l_{\rm cr}^{\rm SK}=3$ found by SK, 
who have used catastrophic synchrotron losses, we have to take the following 
steps:
\begin{enumerate}
 \item 
The case of synchrotron cooling degenerates to the `catastrophic losses case' when the threshold energy for
 absorption $2/\egamma$ equals the maximum soft photon energy $\xmax$. \\
\item
In SK the volume element is defined as $V=\pi R^3$, whereas we used $V=4/3 \pi R^3$. 
Thus, the replacement $l^{\rm SK}_{\rm cr} \rightarrow \frac{4}{3} l_{\rm cr}^{\rm SK}$ should be made. \\
\item
Finally, in SK the luminosity compactness is not given by the conventional definition, which we adopted, but is multiplied by the energy $\egamma$.
\end{enumerate}
Thus, our result should be compared to $\frac{4 l_{\rm cr}^{\rm SK}}{3\eg}$. 
The critical compactness as a function of $\eg$ for two values of the magnetic field is shown in Fig.~\ref{lcr1}.
The two crosses mark the values of SK, which completely agree with ours in the limit of catastrophic losses.

We note also that eq. (\ref{lcr0}), which was obtained by making robust calculations, also 
agrees with the eq. (\ref{lcrit_delta}) in the limit
of catastrophic losses. 

\subsubsection{Step function approximation for the $\sgg$ cross section}
In this paragraph we analytically find the critical luminosity compactness after adopting a more realistic expression
for the annihilation cross section:
\eqb
\sgg=\sth \sigma_0 \frac{\Theta(x\egamma-2)}{x\egamma},
\eqe
where the normalization used is $\sigma_0=4/3$.
The initial equations (\ref{gammaray})-(\ref{elec}) keep the same form. Only the operator of
photon losses becomes 
\eqb
\cal{L}^{\gamma}_{\gamma \gamma}
& = & n(2 \gamma)\int dx \sigma_0 \frac{\Theta(x \eg-2)}{x\eg}n_0'(x) \cdot\eqe
Following the same steps as those described in \S 2.2.1, we find for $s=0$
\eqb
\label{elec_theta1}
\gamma^2 n_e'(\gamma) = C-A\int_{\gcr}^{\gmax} d\gamma \gamma^{-2}n_e'(\gamma),
\eqe
where $A=\frac{2\bar{n} \sigma_0}{b^2 \eg}$ and $\gmax=\eg/2$.
Equation (\ref{elec_theta1}) can be solved iteratively when writing $n_e'(\gamma)$ 
as a sum of approximations, i.e., $n_e'(\gamma)=n^{(0)}+n^{(1)}+\cdots$. 
Assuming for the first approximation the form $n^{(0)}=C\gamma^{-2}$,
 we find the other terms of the series expansion:
\eqb
\label{approx}
n^{(1)} & = & C\gamma^{-2}\lambda \nonumber \\
n^{(2)} & = & C\gamma^{-2} \lambda^2,\\
 \vdots 
\eqe
where $\lambda=\frac{A}{3}\left(\frac{1}{\gmax^3}-\frac{1}{\gcr^3}\right)$ and $|\lambda|<1$
for typical values of the parameters.
Thus, the series of approximations 
converges and $n_e'$ can be found in closed form:
\eqb
\label{elec_theta2}
n_e'(\gamma) & = & \frac{C}{\gamma^2} \sum_{n=1}^{\infty} \lambda^{n-1}= \frac{C}{\gamma^2 (1-\lambda)}\cdot 
\eqe
The luminosity in soft and hard photons can then be found
\eqb
l_{\gamma}^{s '} = \frac{4\lB }{3}\frac{C \eg}{2(1-\lambda)}
\eqe
\eqb
l_{\gamma}^{h '} = \frac{2 \lB C \eg^2}{3} \frac{|\lambda|}{1-\lambda}\cdot
\eqe
The critical $\gamma$-ray compactness is given by
\eqb
\label{lcr_theta}
\lcr= \frac{b^2\eg^2}{2 \sigma_0}\left[\left(\frac{b \eg}{2}\right)^{3/2}-\frac{8}{\eg^3}\right]^{-1} \cdot
\eqe
\begin{figure}[h!]
 \centering
\includegraphics[width=8.5cm, height=8cm]{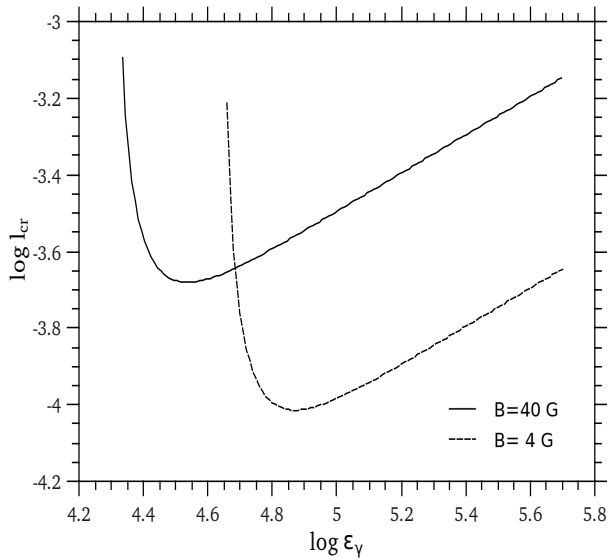}
\caption{Log-log plot of the critical luminosity compactness $\lcr$ as a function of the injected $\gamma$-ray
 photon energy $\egamma$, when the $\Theta$-function approximation for the cross section $\sgg$ was used
 for two values of the magnetic field strength: $B=40$ G (solid line) and $B=4$ G (dashed line). }
\label{lcr2}
\end{figure}
 Figure \ref{lcr2} shows $\lcr$ as a function of $\eg$ for two values of the magnetic field. When the 
threshold energy for absorption
$2/\eg$ is much lower than the maximum energy of the soft synchrotron photons $b\eg^2/4$, the critical compactness as a function of $\eg$
behaves as for the $\delta$-function cross section. However, when the two energies become comparable, $\lcr$ abruptly increases, because
the number of soft photons serving as targets for the absorption significantly decreases. This effect
can be seen in Fig.~\ref{lcr2} as a change in the curvature of the curves.

 \section{Numerical approach}

To numerically investigate the properties of quenching,
one needs to solve again the system of eqns (5)-(7) 
augmented to include more physical processes. 
As in the numerical code, there is no need to treat the time-evolution of
soft photons and $\gamma-$rays through separate
equations, the system can be written:

\eqb
{{\partial\nelec(\gamma,t)}\over{\partial t}}+
{{\nelec}\over{\teesc}}= \cal{Q}^{\rm e}_{\gamma\gamma}
+\cal{L}^{\rm e}_{\rm syn}
+\cal{L}^{\rm e}_{\rm ics}
\eqe
and
\eqb
{{\partial\nphot(x,t)}\over{\partial t}}+
{{\nphot}\over{\tgesc}}= \cal{L}_{\gamma\gamma}^\gamma
+\cal{Q}_{\rm syn}^\gamma
+\cal{Q}_{\rm ics}^\gamma
+\cal{L}_{\rm ssa}^\gamma
+\cal{Q}_{\rm inj}^\gamma,
\eqe
where $\nelec$ and $\nphot$ are the differential electron
and photon number density, respectively, normalized as
in \S2. Here we considered the following processes:
(i) Photon-photon
pair production, which acts as a source term for electrons
($\cal{Q}^{\rm e}_{\gamma\gamma}$) and a sink term for photons
($\cal{L}_{\gamma\gamma}^\gamma$); (ii) synchrotron radiation,
which acts as a loss term for electrons ($\cal{L}^{\rm e}_{\rm syn}$)
and a source term for photons ($\cal{Q}_{\rm syn}^\gamma$);
(iii) synchrotron self-absorption,
which acts as a loss term for photons (${\cal{L}}^{\gamma}_{\rm ssa}$)
and (iv) inverse Compton scattering, 
which acts as a loss term for electrons ($\cal{L}^{\rm e}_{\rm ics}$)
and a source term for photons ($\cal{Q}_{\rm ics}^\gamma$).
In addition to the above, we assume that $\gamma-$rays are
injected into the source through the term ($\cal{Q}_{\rm inj}^\gamma$).
The functional forms of the various rates have been presented
elsewhere -- \cite{mastkirk97} and \cite{petromast09} . The photons are assumed to escape
the source in one crossing time, therefore $\tgesc=R/c$.

The parameters of the problem are completely specified once
the values of radius $R$ and of the magnetic field $B$ are set
and the injection rate $\cal{Q}_{\rm inj}^\gamma$ is specified.
Because  $\cal{Q}_{\rm inj}^\gamma$ is readily related to the injected
luminosity $L_\gamma^{\rm inj}$ through the relation
\eqb
\label{normalization}
L_\gamma^{\rm inj}=(\melec c^2)^2 V \int dx \ x \
\cal{Q}^{\gamma}_{\rm inj},
\eqe
where $V$ is the volume of the source,
the injection rate is specified once the injection
compactness $\lginj$
and the functional dependence of 
$\cal{Q}_{\rm inj}^\gamma$ on $\eg$ is set.
The electron physical escape timescale from the source
$\teesc$ is another free parameter which, however, is not
important in our case. Thus, we will fix it
at value $\teesc=\tgesc=R/c$.
 The final settings are the initial  conditions for the electron 
and photon number densities. Because we are investigating 
the spontaneous growth of pairs and synchrotron photons,
we assume that at $t=0$ there are no electrons or photons in the source, so we set
$\nelec(\gamma,0)=\nphot(x,0)=0$. 

\subsection{Monoenergetic $\gamma-$ray injection}

As a first example we study the case where the $\gamma-$rays injected
are monoenergetic at an energy $\eg$.
The expected trivial solution of the kinetic equations is that because
 there are no soft photons in the source, the injected $\gamma-$rays
will freely escape and their spectrum upon escape will simply
be their production one. Electrons and photons at energies 
different from $\eg$ will remain at their initial values, i.e.,
$\nelec(\gamma,t)=\nphot(x,t)=0$.

\begin{figure}
\centering
\includegraphics[width=8.4cm,height=8.1cm]{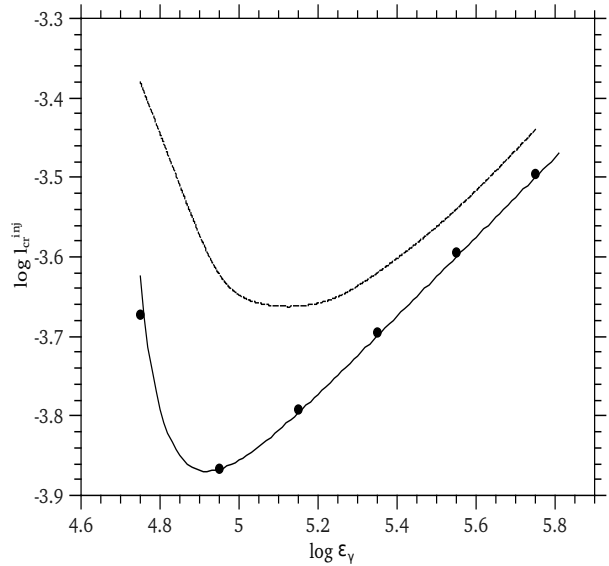}
\caption{Critical $\gamma$-ray
compactness as a function of $\eg$ when (i) $\sgg$ and synchrotron emissivity are approximated by a $\Theta$- and a $\delta$- function
respectively and (ii) when for both quantities the full expressions are used. For the first case the numerical
result (points) and the analytical one given by eq.(\ref{lcr_theta}) (solid line) are shown. 
The full problem can only be treated numerically and the corresponding result is shown with a dashed line. The size of the source
is $R=3\times 10^{16}$ cm and the magnetic field strength $B=3.57$ G.}
\label{compare1} 
\end{figure}

Before numerically examining the stability of the above system using the full expressions for
the photon-photon annihilation cross section and the emissivities, we 
perform a test by comparing the analytical
result given by eq. (\ref{lcr_theta}) with the respective numerical. The two solutions
completely agree as shown in Fig.~\ref{compare1}. In the same figure we also plot
the critical $\gamma$-ray compactness found numerically when all quantities used were given by their full 
expressions. The qualitative behaviour of the full problem solution is quite well reproduced by the analytical one
obtained while using the $\Theta$- and $\delta$- approximations for the cross section and synchrotron emissivity
respectively. This allows us to use the analytical expression given by eq. (\ref{lcr_theta}) for
the full problem as well, when investigating the properties of quenching at least qualitatively.
From this figure it becomes apparent that $\gamma$-ray quenching can affect all photons with energies
that satisfy the feedback condition. Starting around this energy and moving progressively to higher ones, we find that
the critical compactness first sharply decreases until it reaches a minimum and then starts increasing again,
 as $\lcr \propto \eg^{1/2}$ asymptotically. 
The exact value of the critical compactness also depends on the strength of the magnetic field. 
For a higher B-field the required minimum energy $\egamma$ decreases, which can be easily deduced by eq.(\ref{bcrit}).
Note also that because of the marginal relation (\ref{lcr_theta}), the required critical injected
compactness becomes higher. Thus, by increasing $B$, the
corresponding curve is shifted towards lower $\gamma-$ ray energies and higher critical 
compactnesses, keeping its shape (see e.g. Fig.~\ref{lcr2}).

\begin{figure}[h]
\centering
\resizebox{\hsize}{!}{\includegraphics{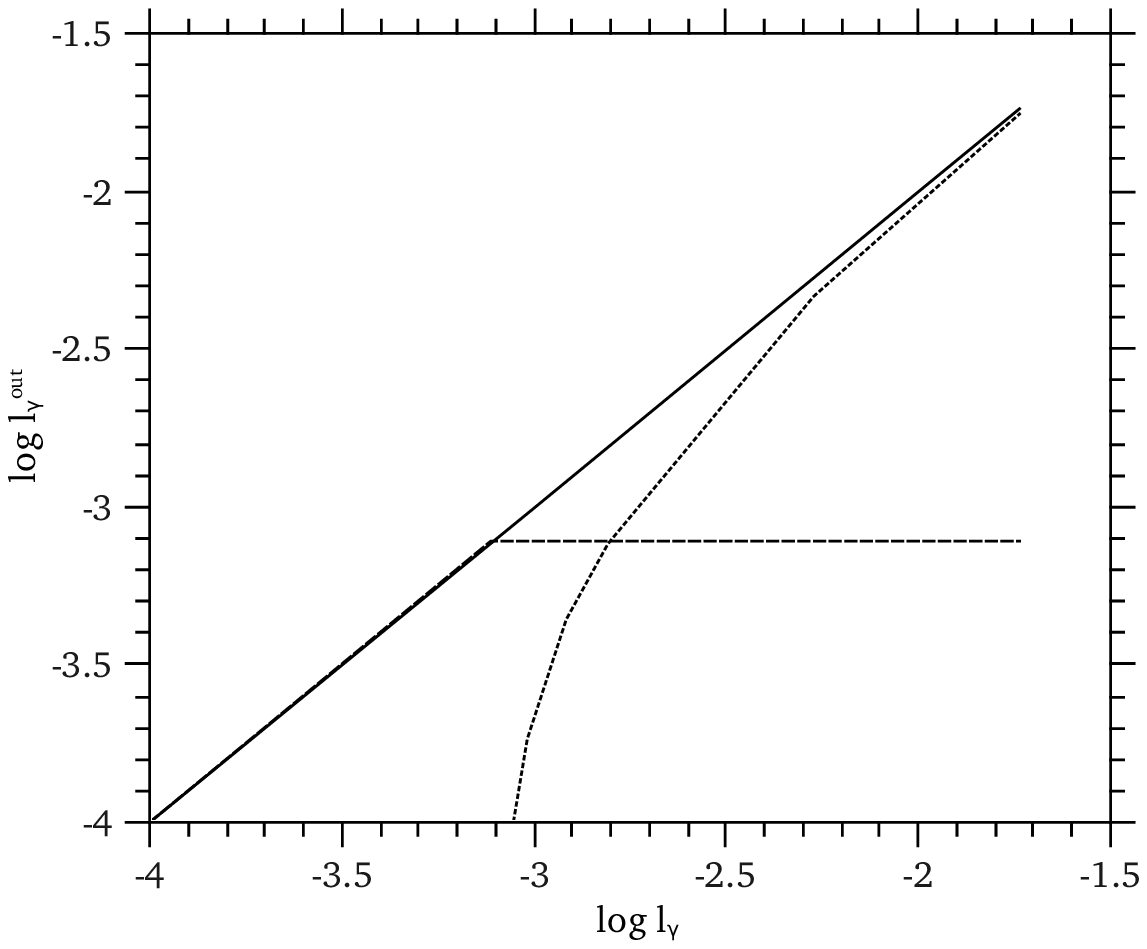}}
\caption{
Plot of the compactness of emerging hard 
$\lgh$ (dashed), soft $\lgs$ (dotted), and total
radiated luminosity $\lgo$ (full line) as a function 
of the injected compactness $\lginj$. 
The $\gamma-$rays  
are assumed to be monoenergetic, injected at energy
$\eg=2.3\times 10^4$.
The size of the source is 
$R=3\times 10^{16}$ cm, while the magnetic field strength
is $B=40$ G. The emerging soft radiation is a result
of the self-quenching of the system.} 
 \label{compact_soft_hard}
\end{figure}

Figure \ref{compact_soft_hard} depicts the results of a series of runs where we kept
$\eg$ constant and varied $\lginj$. 
For sufficiently low values of the latter parameter, all $\gamma-$rays indeed
escape from the system without any impedance.
If we call $\lgh$ the compactness of the
outgoing radiation at energy $\eg$, then obviously,
in this case,  $\lgo=\lgh=\lginj$,
where $\lgo$ is the compactness of the outgoing luminosity. 
However, above  a critical value
of $\lginj$, which we will call $\lginjcr$, 
the loop described above starts operating,
transferring the $\gamma-$ray luminosity to softer radiation.
Thus, for $\lginj>\lginjcr$, we get $\lgo=\lgs+\lgh=\lgs+\lginjcr$
because $\lgh$ saturates -- by $\lgs$ we denote the compactness
of the reprocessed luminosity, which appears at lower energies.
This can be seen in Fig.~\ref{compact_soft_hard}, where the dashed line curve 
represents $\lgh$, the dotted line $\lgs$ and the full line
$\lgo$, which is always equal to $\lginj$ because there are no sinks
of photons present. For the particular example shown in Fig.~\ref{compact_soft_hard}, the 
values of the other parameters used are, 
$R=3\times 10^{16}$ cm, $B=40$ G and $\eg=2.3\times 10^4$.

\begin{figure}[h]
 \centering
\resizebox{\hsize}{!}{\includegraphics{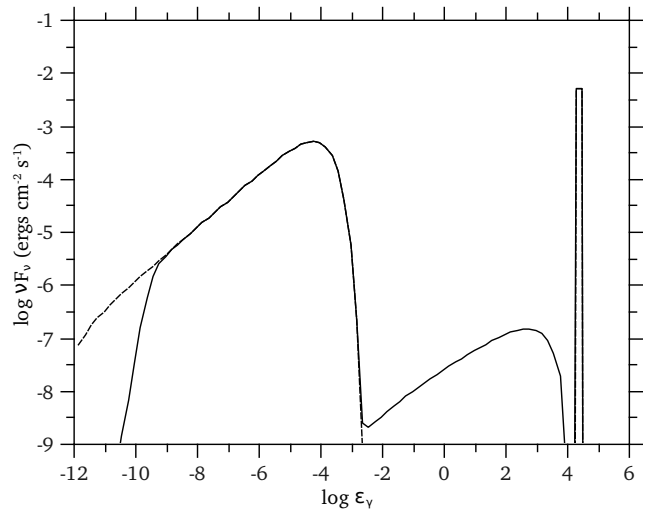}}
\caption{Multiwavelength spectrum for a choice of
$\lginj$ such that $\lgs=\lgh$. The other parameters
are the same as in Fig.~\ref{compact_soft_hard}.
 Full line corresponds to the
case where all relevant radiative processes were taken
into account, while the dashed line corresponds to the
case where only the two `core' processes were
considered, i.e., synchrotron and $\ggabs$ interactions.} 
\label{MW}
\end{figure}

Figure \ref{MW} shows the multiwavelength (MW) spectrum for the
particular value of $\lginj$, which results in equipartition
between soft and hard luminosities, i.e., $\lgs=\lgh$.
While the hard luminosity is simply the escaping
monoenergetic radiation at $\eg$, the soft spectrum 
mainly consists of the synchrotron component 
produced from the secondary pairs, which were injected
at energies $\gamma_p=\gamma_e=\eg/2$. 
The peak of this component is at $\epsilon^{s}_{\rm pk}=b\gamma^2$,
while it has a spectral index of $-1/2$ (or $+1/2$ in
$\nu F_{\nu}$ units) as a result of the electron cooling
to a $\gamma^{-2}$ distribution below the energy of injection. 
There is also a 
synchrotron self-Compton (ssc) component, 
which is greatly suppressed by Klein-Nishina effects.
As can be seen from the figure, the corrections 
introduced from the inclusion of inverse Compton scattering
and synchrotron self-absorption are negligible,
at least for our set of parameters.

\begin{figure}[h!]
\centering
\resizebox{\hsize}{!}{\includegraphics{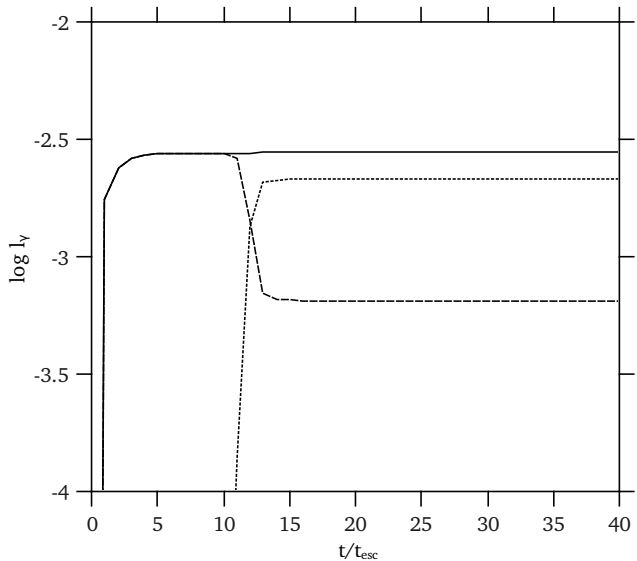}} 
\caption{Time evolution of the soft $l_{\gamma}^{s}$ (dotted line), hard $l_{\gamma}^{h}$ (dashed line)
and total $l_{\gamma}$ (solid line) compactness for the same set of parameters
as in Fig.~\ref{MW}.}
\label{lgt}
\end{figure}

Figure \ref{lgt} shows the time evolution of the different luminosity compactnesses for the example case presented
in Fig.~\ref{MW}. Although the dynamical, non-linear system of equations describing the system 
(see for example system of eqns. (\ref{systemfull})) is of third order, which in principle 
can lead to unstable solutions or/and `chaotic' behaviour (SK), the system quickly reaches 
a stable state after a few crossing
times. Figure \ref{lgt} shows that
although initially absent, a soft photon population builds up, which leads
to a decrease of $l_{\gamma}^h$,
because $\gamma$-rays are being absorbed by the soft photons.

\subsection {Power-law injection}

As a next step we investigate the effects of quenching
for a power-law injection of $\gamma-$rays,
i.e., when the injection rate can be written as
\eqb
Q_\gamma=Q_{0} \eg^{-\alpha}\Theta(\eg-\egmn)\Theta(\egmx-\eg),
\eqe
where $Q_0$ is a normalization constant depending on the value of $\alpha$,
the low- and high- energy cutoffs of the distribution.

Before proceeding to more quantitative calculations,
we remark that obviously, for a
given $B$ the feedback criterion for quenching to start
will be that of eq. (\ref{bcrit}) where $\eg$ should be replaced
by $\egmx$. Assuming that this condition holds
and that the compactness of the $\gamma-$rays 
is above the critical one, the production
of synchrotron photons with energy $\esynmx=b\egmx^2/4$
is guaranteed. This means that $\gamma-$rays with
energy $\eg \gtrsim\egmneff$ should interact in $\ggabs$ interactions,
where 
\eqb
\label{gmineff}
\egmneff={{2}/{\esynmx}}={{8}/{b\egmx^2}}.
\eqe
Thus, depending on the relation between 
$\egmneff$ and $\egmn$ we can distinguish two cases

(i) $\egmn < \egmneff$: In this case quenching
will affect $\gamma-$rays from $\egmx$ down to $\egmneff$.
For $\egamma\le\egmneff$ the spectrum will be unaffected,
i.e., it will be the same as the one at injection.

(ii) $\egmneff < \egmn$: In this case quenching
will affect the whole $\gamma-$ray spectrum.

\begin{figure}[h]
 \centering
\resizebox{\hsize}{!}{\includegraphics{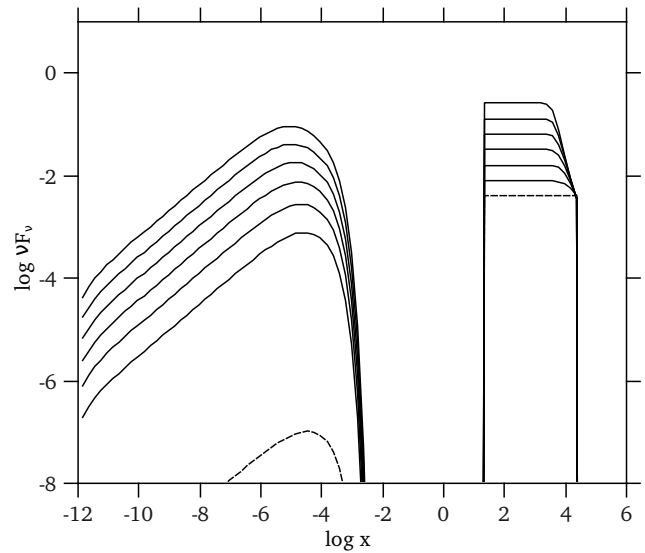}}
\caption{MW spectra for a power-law injection of
$\gamma-$rays of the form $Q_\gamma=Q_0\eg^{-2}$ 
for $\egmn<\eg<\egmx$
with $\egmn=23$ and $\egmx=2.3\times 10^4$
for various values of the normalization $Q_0$.
This starts from $Q_0=4\times 10^{-3}$ and then exceeds
 its previous value by a factor of 2. 
The dashed line curves represent the runs without quenching.
The source parameters are 
$B=40$ G and $R=3\times 10^{16}$ cm.}
\label{partly_quench}
\end{figure}

\begin{figure}[h]
 \centering
\resizebox{\hsize}{!}{\includegraphics{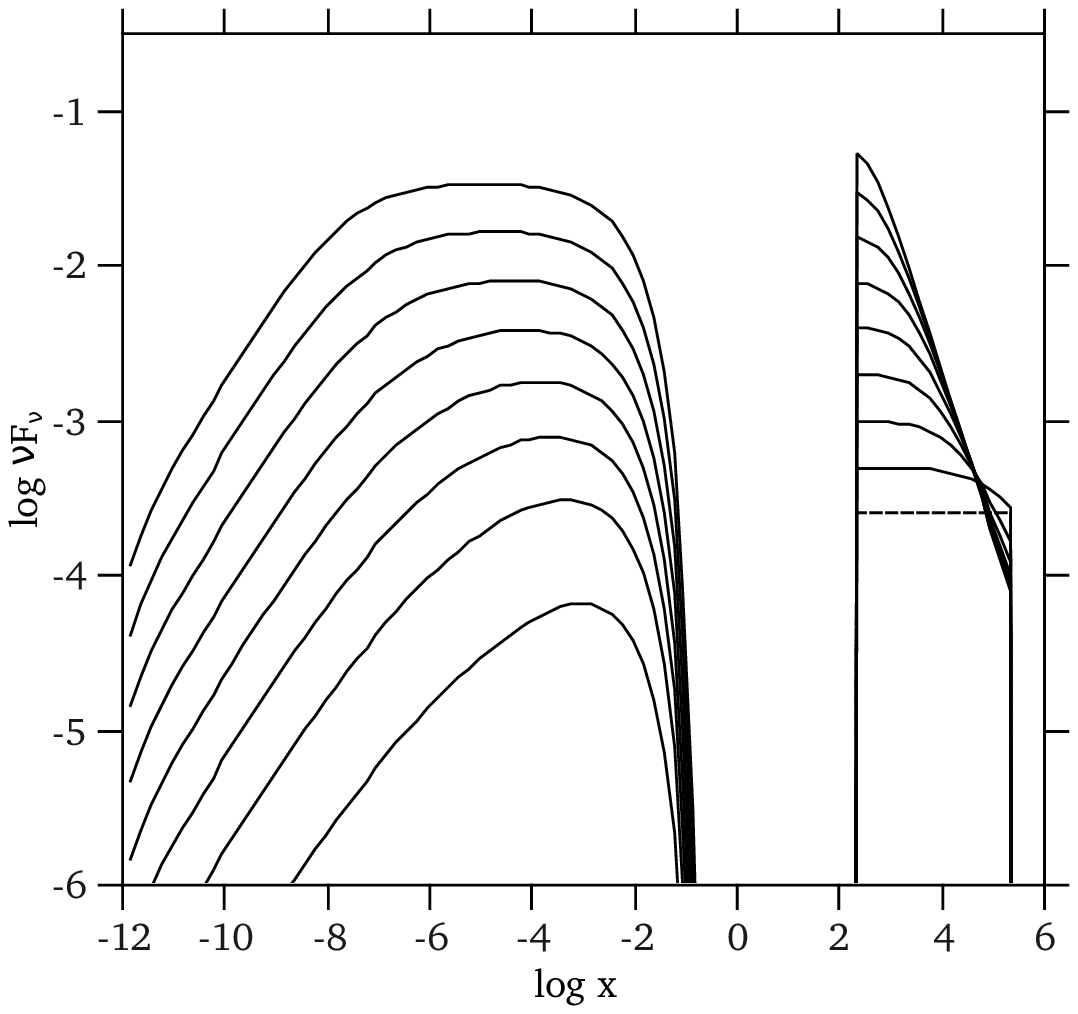}}
\caption{Same as in Fig.~\ref{partly_quench}
with $\egmn=2.3\times 10^2$ and $\egmx=2.3\times 10^5$
with $Q_0$ starting from $Q_0=2.5\times 10^{-4}$.}
\label{all_quench}
\end{figure}

The above two cases are exemplified in Figures \ref{partly_quench} and \ref{all_quench}.
In both figures we assumed that $\gamma-$rays with a
power-law form of slope $\alpha=2$ are injected in a source
of radius $R=3\times 10^{16}$ cm and magnetic field
strength $B=40$~G. The only difference between the two cases
is in the lower and
upper limits, which are $\egmn=23$ and $\egmx=2.3\times 10^{4}$
for the case shown in Fig.~\ref{partly_quench} and 
$\egmn=230$ and $\egmx=2.3\times 10^{5}$
for the case shown in Fig.~\ref{all_quench}. 

In the case of Fig. ~\ref{partly_quench} we have $\egmn<\egmneff=1.7\times 10^4$ 
and, as can be seen from the figure, the power-law breaks
above $\egmneff$ because of quenching -- indeed,
we find that the value of $\egmneff$ is in reality lower by a factor
of 2  than the one given in eq. (\ref{gmineff}), because in the numerical
calculations shown here we use the full expressions for the
synchrotron emissivity and the cross-section for $\ggabs$ 
and not the $\delta-$function approximations implied in 
the derivation of the above relation. It is interesting to note
that as the $\gamma-$ray luminosity increases and thus
the quenching becomes more effective, the break between
the unabsorbed and absorbed part of the power-law becomes
more abrupt.

In the case of Fig. ~\ref{all_quench}, on the other hand, the choice of
the parameters were such as to make $\egmneff$ comparable
to $\egmn$. Thus, according to the qualitative analysis 
presented above, quenching should affect the whole 
$\gamma-$ray distribution. Indeed, as can be deduced
from the figure, as the luminosity increases, absorption
is moving progressively to lower $\gamma-$ray
energies, until it affects the whole distribution.

\begin{figure}
 \centering
\resizebox{\hsize}{!}{\includegraphics{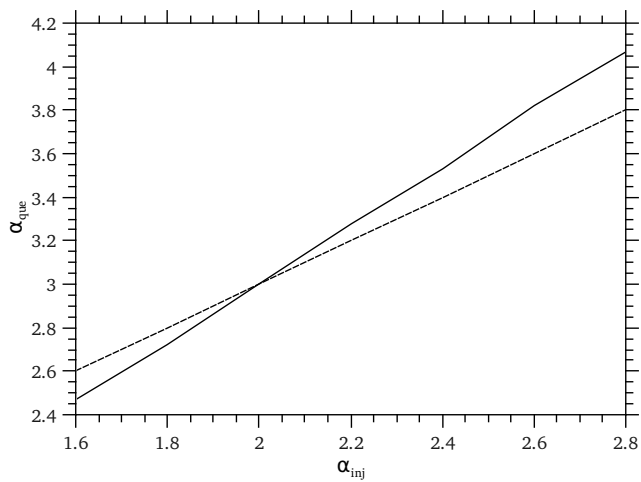}}
\caption{Slope of the quenched $\gamma$-ray spectrum as a function
of the injected spectrum's slope (solid line). To facilitate a comparison 
the line $\alpha_{\rm inj}+1$ is also shown (dashed line).}
\label{ainj}
\end{figure}

We also investigated the relation between the slope of the injected ($\alpha_{\rm inj}$) and the `quenched' ($\alpha_{\rm que}$)
$\gamma$-ray spectrum. For that reason,
 we ran the code for different $\alpha_{\rm inj}$, choosing each time suitable 
values of the source parameters to enable quenching. 
Figure~\ref{ainj} shows the result of the runs (solid line) and the slope $\alpha_{\rm inj}+1$
(dashed line) for comparison reasons. We find that $\alpha_{\rm que} \propto \alpha_{\rm inj}^{1.34}$.
Thus, quenching becomes more pronounced as the injected spectrum
gets steeper. 

\section{Application to the one-zone homogeneous
models: The 3C 279 case}

The above has a straightforward application to the 
so-called one-zone homogeneous models, which are
customarily used to fit the MW spectra of Active Galactic Nuclei (AGN)
\citep{boettcher09, aleksic11}. According to these models,
the `source' contains
electrons and/or protons, which radiate through various
processes and form a spectrum that can then be
fitted to the observed data by varying certain key 
parameters -- for a detailed description see \cite{mastkirk97}
and \cite{konopelko03}. 

Alternatively, one can assume that the $\gamma-$rays
are produced by some mechanism and investigate the
allowed parameter space that allows the escape of these
$\gamma$-rays. To perform this analysis, one
does not require the existence of other softer photons
that act as targets, but one simply has to look for
the threshold conditions for photon quenching. In other
words, the relevant question one has to ask is:
 {\sl{Assume a spherical source of radius $R$
containing a magnetic field $B$ and producing a 
$\gamma-$ray spectrum that is measured at Earth 
with a flux $F_\gamma(\epsilon_\gamma)$. If the source
is moving with a Doppler factor $\delta$ with respect to us,
what is the  parameter space of $R$, $B$, and $\delta$
that allows the $\gamma-$rays to escape?}}
As we will show, this can give very robust lower limits
for the Doppler factor $\delta$.

The luminous quasar 3C 279 is a good example.
A recent comprehensive
review of observations can be found in \cite{boettcher09}.
Here we will mainly focus on the 2006 campaign, which
discovered the source at VHE $\gamma$-rays, showing a high TeV flux \citep{albert08}, while 
the X-rays were at a much lower level.

To apply the ideas discussed in the previous section,
we follow the procedure below:
 
We assume a spherical source of radius $R$ 
moving with Doppler
factor $\delta$ with respect to us
and containing a magnetic field of strength $B$.
We assume that $\gamma-$rays are produced by some unspecified
mechanism inside the source and, upon escape, they form the observed 
$\gamma-$ray spectrum. If $\lginj$ is the compactness of the $\gamma-$rays
(c.f. eq. \ref{lgg}), we can relate it to the observed integrated
flux $F^{\rm int}_\gamma$ by the relation
\eqb
\lginj={{3\sigma_T F^{\rm int}_{\gamma} D_L^2}\over{\delta^4 R\melec c^3}}
\eqe
where $D_L=3.08$ Gpc is the luminosity distance to the source. Here we
 used a cosmology with $\Omega_m=0.3$, $\Omega_\Lambda=0.7$
and $H_0=70$ km s$^{-1}$ Mpc$^{-1}$. 

To model the TeV emission, we performed a $\chi^2$
fit to the MAGIC data, corrected 
for intergalactic $\gamma \gamma$ absorption \citep{boettcher09}, of the form
$\nu F_\nu=A\nu^{-\alpha}$ for 
$\numn<\nu<\numx$ where
$A=1.3\times 10^{31}$ Jy$\cdot$Hz, $\alpha=0.7$,
$\numn=2\times 10^{25}$ Hz and
$\numx= 10^{26}$ Hz.
Then we injected  $\gamma-$rays 
in the code of the form
$Q_\gamma=Q_0\eg^{-\alpha}$
for $\egmn<\eg<\egmx$ 
where the normalization is set by
condition (\ref{normalization}) with the integration
taking place between

$\egmn=h\numn/(\delta\melec c^2)$ and
$\egmx=h\numx/(\delta\melec c^2)$. 

\begin{figure}[h]
 \centering
\resizebox{\hsize}{!}{\includegraphics{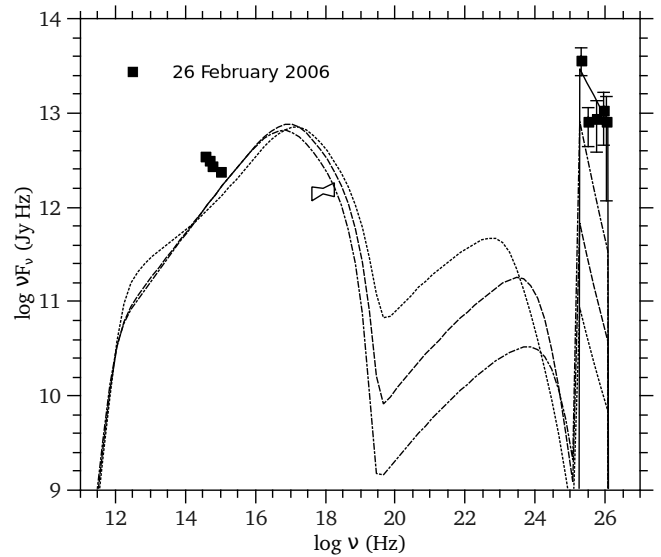}}
\caption
{MW spectra of 3C 279 for 
$R=10^{16}$ cm, $B=40$ G and $\delta=10^{0.8}$ (dotted line),
$10^{1.0}$ (dashed line),
$10^{1.2}$  (dot-dashed line) and
$10^{1.4}$ (solid line).
In all cases the $\gamma$-ray injection would have produced
the solid line in the absence of quenching. 
Black squares and bowtie represent the observational data from
February 2006.}
\label{mw_doppler}
\end{figure}

\begin{figure}[h]
\centering
\resizebox{\hsize}{!}{\includegraphics{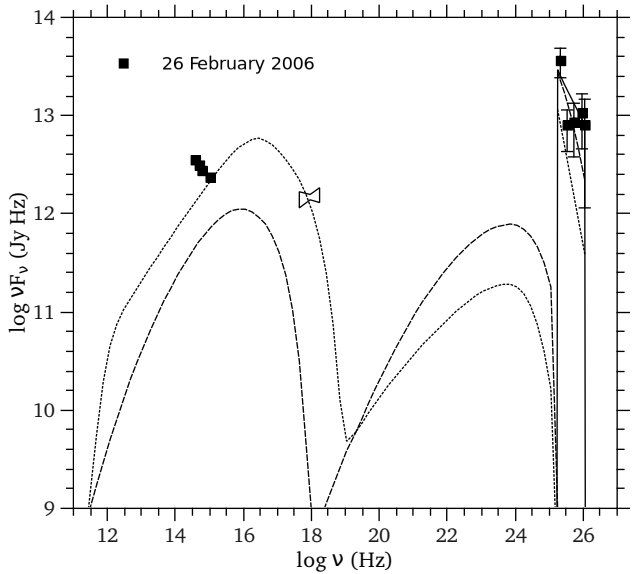}}
\caption{MW spectra of 3C 279 for 
$R=10^{16}$ cm,  $\delta=10^{1.2}$ 
and $B=$.16~G (solid line), 1.6~G (dashed line) 
and 16~G (dotted line).
In all cases the $\gamma$-ray injection would have produced
the solid line in the absence of quenching. 
Black squares and bowtie represent the observational data from
February 2006.}
\label{mw_b}
\end{figure}

Fig.~\ref{mw_doppler} depicts the resulting MW spectra for 
$R=10^{16}$ cm, $B=40$ G and $\delta$
varying between $10^{0.8}$ to $10^{1.4}$
with increments of 0.2 in the exponent. 
Only the run with the highest value of $\delta$
can fit the TeV data. The other three, even though
they have parameters that could also fit the data,
they do not because of quenching. It is important 
to repeat at this point that the $\gamma-$ray luminosity produced
in these cases is the maximum possible that the system
can radiate for the particular choice of $R$, $B$,
and $\delta$. Any effort to increase the $\gamma-$ray
flux by e.g. increasing $\lginj$ would only result in an
increase of the reprocessed radiation.

Fig.~\ref{mw_b} depicts a series of similar runs with 
$R=10^{16}$ cm, $\delta=10^{1.2}$ and $B=.16$~G 
(solid line), $B=1.6$~G (dashed line) and
$B=16$~G (dotted line). While for low $B$ values there is no quenching,
this begins to appear at higher values of the magnetic field.
Note, however, that the value of $B=1.6$~G produces
an acceptable fit, because the soft reprocessed
radiation does not violate any observable limit (most notably
the X-rays) and it can produce a marginally acceptable fit to
TeV $\gamma-$rays. Based on this, one can define a limiting
value of the magnetic field
$\bquemx(R,\delta)$ with the property that if $B>\bquemx$,
the quenching does not allow the internally produced $\gamma-$ray 
luminosity to escape. Therefore $\bquemx$ is the highest allowed
value that the $B-$field can take for the observed flux
of TeV $\gamma-$rays.

\begin{figure}[h]
\resizebox{\hsize}{!}{\includegraphics{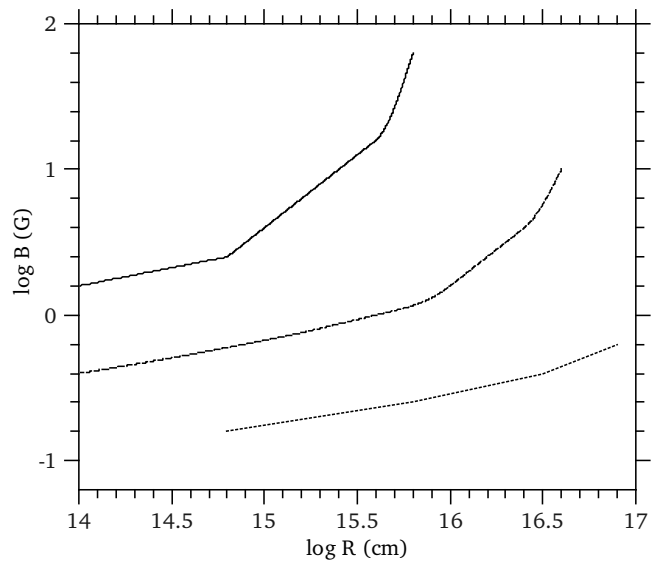}}
\caption{Parameter space of the allowed values of the magnetic
field B as a function of the size of the source R for
Doppler factors of $\delta=$10 (dotted line), 15.8 (dashed
line), and 25 (solid line).  The space above a line
for a certain $\delta$ corresponds to combinations of R and B
that cannot successfully fit the TeV observations of 3C 279.}  
\label{Blocus}
\end{figure}

Figure \ref{Blocus} depicts the locus $\bquemx$
as a function of
the size of the source $R$ for various Doppler factors $\delta$.
Each line separates the parameter space into two
regimes: Because of the quenching, a choice of $R$ and $B$
 above the locus produces 
a far too low $\gamma-$ray flux that
cannot fit the TeV observations.
For example, if one assumes that $\delta=10$
and $R=2.5\times 10^{16}$~cm -- so that it corresponds
to a variability timescale of 1 day,
 the magnetic field
cannot exceed 0.3 G. However, owing to the steep
dependence of the flux on $\delta$ (c.f. Eqn 40),
the above condition is greatly relaxed for higher values
of $\delta$.

We should also mention that $\bquemx$ not only 
depends on $\delta$ and $R$, but also
on the upper cutoff $\egmx$
of the injected power-law in $\gamma-$rays.
For the calculations  performed here,
we chose $\egmx$ such as to match
the highest observed frequency in the data set,
i.e., $10^{26}$~Hz. If one assumes that 
the internally produced $\gamma-$rays can extend to even
higher energies, then the value of $\bquemx$
is affected as well. Accordingly we found that if we were
to repeat the same calculations with $\egmx$
ten times higher than the value we used,
then quenching would affect the flux more
and the lines of Fig.12 would have to shift upwards
by a factor of 2. Therefore, Fig.12
can be taken as a conservative lower limit 
on $\bquemx$. 

\section{Summary and discussion}

We have examined the effects of photon quenching
on compact $\gamma-$ray sources. Automatic photon quenching, first proposed
by SK, is a non-linear network of two processes that are very 
common in high-energy astrophysics, namely 
photon-photon annihilation and lepton synchrotron radiation. 
We verified
and expanded both analytically and numerically the results
of SK and we showed that this network, when applied to compact
$\gamma-$ray sources, can give  
limits on the Doppler factor, which depend on the source
radius and on the magnetic field strength only.
This is a theoretically derived limit and as such does not
depend 
on the soft photon populations that might be present 
inside the source.

It is important to note that photon quenching can be investigated
only when the radiation transfer problem is solved self-consistently
because it is a non-linear process that involves redistribution
of energy from the $\gamma-$rays to the lower energy parts of
the photon spectrum. For this we used 
a system of kinetic equations that treats in the simplest case
$\gamma-$ray annihilation
 and electron synchrotron radiation.
Using suitable $\delta-$function approximations
for the photon-photon annihilation cross section and 
for the photon synchrotron emissivity, 
we were able to verify the results of SK in the limit where
synchrotron losses can be considered as catastrophic (see Fig.~\ref{lcr1}).
On the other hand, using a more realistic step function
for the cross section, we were able to find a very good agreement
with numerical results, which utilized the 
full photon-photon annihilation cross section.
Furthermore, we extended our numerical study in the case
where $\gamma-$rays are injected in the form of a power-law.
We found that quenching can be responsible for wide
spectral breaks, which become more pronounced as the injected
$\gamma-$rays become steeper (see Fig.~\ref{ainj}).

Quenching has some interesting consequences
when applied to existing $\gamma-$ray sources.
For example, for the 2006 MAGIC 
observations of the quasar 3C279, modelling with
combinations of low values of the Doppler factor $\delta$
and high values of the magnetic field strength $B$ lead
to strong absorption of the injected $\gamma-$rays.
The only way to avoid quenching is either to assume
typical values of $\delta$ (i.e., around 10)
and low values of $B$
or, alternatively, high values for both $\delta$
and $B$ -- see Fig.~\ref{Blocus}.

We restricted our analysis to cases where $\gamma-$rays 
are injected into the source without being too specific in
assigning a particular radiation mechanism for producing 
them. We preferred this
approach because our aim was to calculate the effects of
quenching using simple functional forms for the $\gamma-$rays
and not to model the source in detail. For example, we could
have used proton synchrotron radiation to fit the 3C279
TeV observations. However, this would have involved
extra parameters for the radiating relativistic protons  
and would have complicated the analysis beyond the
level at which we would like to present it here.

In our analytical treatment we ignored inverse Compton scattering,
because its inclusion would have complicated the analysis. 
On the other hand, it was taken
into account in all  our numerical calculations, although it
is greatly suppressed by Klein-Nishina effects (see Fig.5).
This process can still make an impact in cases 
where the soft photon compactness greatly exceeds the
magnetic one (c.f. Eqn. 14). However, this has to be seen
in conjunction with the overall fitting of the MW spectrum
of the source and we avoided presenting it here,
although we took it into account in \S4 when performing
the analysis on quasar 3C279.

 Concluding, quenching has many far-reaching
 implications for the modelling of compact
 $\gamma-$ray sources such as Active Galactic Nuclei
 and Gamma Ray Bursts. Because quenching
operates in an autoregulatory manner by coupling electrons to photons, its
effects on the photon spectra cannot be manifested if the modelling
 is done by assuming an \textit{ad-hoc} radiating
electron and/or proton distribution. This could lead to parameter choices
that lie within the quenching `forbidden' parameter space, making
 them in essence invalid.

On the other hand, if taken fully into account, quenching can give
 robust limits for important source parameters such as
 its size, magnetic field, and the Doppler factor for a
 given $\gamma-$ray flux. In the case of 3C279 for example, we found that 
  to avoid quenching, the source should have a large
 Doppler factor ($\delta \gtrsim 10$) whenever high $B$ values
 are adopted. Similar constraints could be found for other AGNs with spectra
 extending to even higher energies ($\sim 1-10$ TeV). In these
 cases the conditions for quenching are more stringent, because they 
critically depend on the maximum energy of the $\gamma$-ray photons and are
 certainly worth investigating further.

{\sl{Acknowledgments}}: We would like to thank
Dr. John Kirk for useful comments and discussion.
This research has been co-financed by the
European Union (European Social Fund - ESF) and Greek national funds through
the operational Program `Education and Lifelong Learning' of NSRF - Research 
Funding Program: Heracleitus II.

\bibliographystyle{aa}
\bibliography{16763}
\end{document}